# Android and Wireless data-extraction using Wi-Fi

Bert Busstra
School of Computer Science & Informatics
University College Dublin
Dublin 4, Ireland
bert.busstra@ucd.ie

N-A. Le-Khac, M-Tahar Kechadi
School of Computer Science & Informatics
University College Dublin
Dublin 4, Ireland
{an.lekhac, tahar.kechadi}@ucd.ie

*Abstract*—Today, mobile phones are very popular, fast growing technology. Mobile phones of the present day are more and more like small computers. The so-called "smartphones" contain a wealth of information each. This information has been proven to be very useful in crime investigations, because relevant evidence can be found in data retrieved from mobile phones used by criminals. In traditional methods, the data from mobile phones can be extracted using an USB-cable. However, for some reason this USB-cable connection cannot be made, the data should be extracted in an alternative way. In this paper, we study the possibility of extracting data from mobile devices using a Wi-Fi connection. We describe our approach on mobile devices containing the Android Operating System. Through our experiments, we also give recommendation on which application and protocol can be used best to retrieve data.

*Index Terms*—Android, Wireless data extraction, network protocols, SSHDroid, rsync.

## I. INTRODUCTION

Today, almost every citizen has a mobile phone. According to Central Bureau of Statistics (CBS) (Statistics Netherlands) Mobile Internet use continues to grow. It also states in a press release that: (i) 60% internet users go online with mobiles; (ii) Young people use mobiles most; (iii) Mobile phone most popular; (iv) Used for communication, news and fun [1]. Basically, mobile devices contain a wealth of information each. This information has been proven to be very useful in crime investigations, because relevant evidence can be found in data retrieved from mobile phones used by criminals. For instance evidence might be extracted from: Address books, SMS/MMS, Call-history, Twitter, web browsing cache, E-mail, Social media apps like Facebook, Instant messaging such as Whatsapp, Viber, Skype, etc.

As a consequence, the investigators investigate mobile devices on a regularly basis, because the data found on those devices might contain crime evidence. Besides, one of the major operating systems found on these mobile devices is Android. According to the International Data Corporation (IDC) Worldwide Quarterly Mobile Phone Tracker, total Android smartphone shipments worldwide reached 136.0 million units, accounting for 75.0% of the 181.1 million smartphones shipped in third quarter of 2012 [2]. It is no surprise that most mobile devices onto which digital evidence is conducted these days is an Android device. Therefore, we only focus on Android devices in this paper.

To investigate a mobile device an investigator needs to interact with the device directly. The examiner must be sufficiently trained to examine the device. He has to think before he acts, because he must know exactly what effects his actions have on the data, and he must be prepared to prove it. If the examiner did change the data, he must be prepared to explain why it was necessary and what data has changed. For this reason the examiner should document every action taken.

In a traditional way, the data from mobile devices can be extracted using an USB-cable. Most Android devices can be connected to a personal computer by using an USB-connection. Using this USB-connection the content of an Android device can be examined. During forensic investigations, software-packages, like XRY [3] or UFED [4], can be used to acquire data from the device.

Indeed, when a device is "rooted" [5] full access to the file-system can also be gained. In that case more information can be retrieved from the device. Basically, Android rooting is the process which allows the users of Android-devices to gain privileged control (known as "root access") within the Android's subsystem. Besides, Android Debug Bridge (ADB) [6] is a versatile command line tool that lets you communicate with an emulator-instance or connected Android-powered device. Using the ADB tool combined with root-access, the examiner can even make physical dumps from the partitions found on the device, using the UNIX "convert and copy file" commands such as *"dd"*. In this case all information can be retrieved for later analysis.

However, while conducting an investigation on an Android device, it may happen that it is not possible to retrieve data from it using an USB-connection. Reasons for these might be a failing cable or a dirty connector. In these cases an examiner will only investigate the external memory-card if it's available in the device. However, not all devices come with external memory. Sometimes the device only contains internal (in-build) memory. The contents of this memory can normally only be extracted using a USB cable-connection. At that point there should be other methods to acquire data from the in-build memory storage.

In this paper, we investigate the feasibility in making a connection to the Android file-system using wireless networking (Wi-Fi). There is very little research in literature that focus on this approach. There are some tools such as MOBILEdit Forensic [7] which also support to extract data from mobile devices using Wi-Fi. However their approach is not disclosed. In our approach, we aim to study simple, efficient method to use Wi-Fi to extract data from Android mobile devices. Indeed, by performing different experiments with our method, we also try to answer the following questions: (i) Is data extraction from an Android-device possible using Wi-Fi? If so, will it be a good alternative for using an USB-connection with the adb-toolset? (ii) Can we extract all data as we do as with USB, like extracting deleted data? (iii) Which network-protocols and free applications can be used and which is the best? (iv) Will the file-system of the Android target-device remain untouched? (v) Will the file attributes be preserved during file-transfer? (vi) Will the given solution work on different versions of the Android?

The rest of this paper is organised as follows. In Section II, we present the background where we review related work in this context. Section III describes our approach of extracting data using Wi-Fi connection. We show our experiments in Section IV. We also discuss and analysing on testing results in this section. Finally, we conclude in Section V.

II. BACKGROUND

A lot is written about Android devices in general but less on the forensic aspects of retrieving data from the Android devices. On the subject of forensic examination on Android devices, [5] describes several Android Forensic Techniques to acquire data from an Android device by a logical or physical way using an USB cable-connection. Indeed, the author shows that to gain access to the data on the file-system of the Android-device, the developer menu-option: "USB-debugging" must be enabled in the settings-menu of the device. If the access to the settings-menu is blocked, it is nearly impossible to gain access to the devices memory. In most cases an investigator needs unlimited access to the user-interface of the Android-device to adjust some options in the settings-menu. Beside [5] also explains the installation and usage of the Android Software Development Kit (SDK) [8] on the investigating computer.

In [9], author has focused on evaluating the efficiency of software tools that support the extraction of information stored on Android devices. Author also describes the architecture of Android systems as well as methods for data extraction. He chose three software tools to evaluate: Oxygen Forensic Suite 2012 standard [10], MOBILedit Forensic [7] and AFLogical [11]. Among these three tools, only MOBILedit allows to use Wi-Fi to extract data. However, author only mentioned that it can use the wireless connection by installing a small application in the mobile phones to pull the data. There is neither further investigation on this capability nor analysis/discussion on using Wi-Fi connection with this software tool.

In [12], author focuses on the design and implementation of an Android application ("app") that automates the collection of useful data for investigating. However, this approach also uses USB cable-connection to extract relevant data.

In [13], authors introduce a method for acquiring forensic-grade evidence from Android smartphones using open source tools. Authors investigate in cases where the suspect has made use of the smartphone's Wi-Fi or Bluetooth interfaces. However they also use USB cable-connection in their solution.

To the best of my knowledge there has been no research on using Wi-Fi for retrieving data from mobile devices, in particular from Android devices, in the context of a forensic setting. An USB cable-connection looks like the most commonly technique among forensic software that is used for retrieving data from mobile devices. Despite MOBILedit Forensic [5] allows to use Wi-Fi connection to extract data, its protocols are not disclosed and it is moreover not free [14].

III. ADOPTED APPROACH

In this section, we describe our approach for using Wi-Fi to extract data from Android mobile devices. In our method, we firstly investigate on different network protocols which allow us to connect to mobile devices using Wi-Fi and to extract relevant data. Next, as we have to install apps on target devices, we should look at the impact of installing apps on an Android device in the context of preserving artefacts in this device. We also describe our user-interactive script for using commands of extracting files and directories from an Android device using Wi-Fi.

Examining the Android devices, we discovered that there was no available option for a user to remotely connect to the Android file-system using Wi-Fi. When an Android device is purchased it does not come with functionality to access the file-system remotely, to back-up user's data for example. Based on this fact, we concluded that to acquire data from an Android device using Wi-Fi, an app should be installed on the device first to make the wireless communication possible. This app includes network protocols to connect to target device and to retrieve data for further examination.

*A. Protocols*

In our approach, we investigate on different protocols that can be listed as: Rsync [15], HTTP-WebDAV [16], SSH [17], etc. Basically, we look at simple, popular protocols that are integrated in most networking apps.

*1) Rsync:* Rsync can be used to sync data across a network, even in both ways, where it can be instructed to preserve the file attributes. To retrieve data, the Rsync program is needed on the Android target-device. In fact, Rsync needs a "listening" service on the target side; an Rsync-server is required. Rsync-server can be found in networking app such as Server Ultimate [18].

*2) WebDAV:* WebDAV is a Web-based Distributed Authoring and Versioning, "it is a set of extensions to the HTTP protocol which allows users to collaboratively edit and manage files on remote web servers WebDAV uses the HTTP protocol" [16]. Its data can be accessed using a web-browser. Several free WebDAV server-apps can be found on the Google Play app-store. When the WebDAV server-service is started on the Android-target it listens for incoming connections. Then the remote WebDAV-share on the Android target-device can be mounted locally on the computer of the investigator.

*3) SSH:* SSH is a cryptographic network protocol for secure data communication, remote login, remote command execution, and other secure services between two networked devices [17]. With an SSH-service in place, it is also possible to transfer data using the commands SCP, SFTP. With SSHFS, it is possible to mount a remote share using an SSH-connection. Once the share is mounted locally, data can be transferred using rsync or the Linux copy command. There are some SSH server-apps which are available through the Google Play app-store such as SSHDroid [19] and SSHelper [20].

We explore moreover some other protocols such as FTP [21], Samba (SMB) [22] to evaluate their feasibility in setting connection and in extracting data from Android devices. Details of this evaluation can be found in the next section.

*B. Forensic implementations of installing apps on Android target*

As we should install system apps on Android devices so we have to measure the impact of installing software on an Android system, the installation and de-installation of the software can be monitored. Both states can be compared with each other using the program Winmerge [23]. The program can be used to make visible which artefacts were placed on the system during the installation and which are left after the uninstallation of an application.

*C. Extraction process*

Our data-extraction process includes three main steps:
- Installing of server apps integrated with necessary network protocols;
- Connecting Android device to local Wi-Fi networks;
- Retrieving files and directories from Android device using Wi-Fi.

In order to ease the usage of commands for downloading (selected) files and directories from a non-rooted/rooted android device using Wi-Fi, we also create a user-interactive bash script. The script must be run from a Linux-shell or terminal

IV. EVALUATION AND DISCUSSION

We start our experiments with the evaluation on different network protocols mentioned in the previous section. Next, we present our testing environment. We also describe and discuss on testing results.

*A. Protocols*

*1) Rsync:* During my research we discovered that rsync does not work with a non-rooted Android-target-device. It might be caused by the usage of a non-conventional *tcp-port*, which has to be used in a non-rooted environment, rather than the standard 873 port under normal conditions. This is because on non-rooted Android devices, the *tcp-ports* below 1024 may not be used for communication.

Besides, through our tests, we conclude that there is currently no standalone *rsync*-server app to be found in the Google Play app-store, which can be used to remotely access the Android file-system to transfer data locally using Wi-Fi. Indeed, the Android-device must be rooted in order to make the using of *rsync* possible.

*2) WebDAV:* After mounting the remote WebDAV-share on the Android target-device, application *davfs* [24] must be installed as root. The remote file-system of the Android target-device will be mounted read-only on a mount-point defined on the investigating computer. For example:

```
host:~# mount -t davfs 'http://192.168.1.119:8888' /media/davfs/ -o ro
```

Next, *rsync* can be used to fetch data to the computer of the investigator:

```
host:~$ rsync -avz --progress /media/davfs/ /home/investigator/evidence
```

The remote file-system must first be mounted on the investigating computer in order to retrieve the data. It is also possible to use the *"wget"* command to fetch files and directories from a WebDAV directory directly, without mounting the remote-folder first. However, by using *wget*, a lot of index.html files are generated. This might affect the directory where the data is stored in the investigating computer. Indeed, by using *wget*, not all permissions (owner, group, authorization) will be preserved.

*3) SSH:* We tested most of the SSH server-apps which are available through the Google Play app-store. Several SSH-servers need root access to install or to function properly. After thorough experiments, we decided to choose SSHDRoid as a SSH-server. We choose SSHDroid because it is easy to set-up and runs smoothly on rooted and non-rooted Android devices, taking almost no system resources. Indeed, the default password for the user root in SSHDroid (and most other SSH-server packages) is admin. The default port-number to connect on non-rooted devices is: 2222; but this is changed to the standard port: 22 on rooted devices.

Besides, with SSHFS [25], it is possible to mount a remote share using an SSH-connection. Once the share is mounted locally, data can be transferred using *rsync* or the Linux copy command: *"cp"*. In this case, firstly the *ssfs*-package must be installed on the investigating computer. Secondly the SSH-server (e.g. SSHDroid) must be installed on the target-device, which will then act as an SSH-server. Next the Android file-system will be mounted read-only (ro) on the computer of the investigator using *sshfs*. Finally, data is copied from the target-device to the investigating-computer with *rsync*. Also the Linux copy command (*"cp"*) could be used to transfer the data. We conclude that we can use SSHFS, but we still need to mount the remote file-system first.

Indeed, with an SSH-service in place it is also possible to transfer data using the commands: SFTP (Secure-FTP) [27] and SCP (Secure copy) [26]. Despite of using the correct "attributes preserving" parameters, we discovered that both SFTP and SCP did not proper transfer the ownership and permissions of the data.

Basically, the SSH-server can also be used as a "listening" server for *rsync*. Once the SSH-server is installed on the Android target-device, the *rsync*-executable is needed on the target-device. Most of the time, the rsync-executable has to be manually transferred to the target-device because many SSH-server apps from the Google Play app-store do not ship with *rsync*. The *rsync*-executable is not included in SSHDroid app too. We should manually install it. An *rsync*-executable for the Android platform is freely available on the internet.

On a non-rooted environment the application, SSHelper would be an alternative of SSHDroid. The advantage of SSHelper is it includes an *rsync*-executable. However, the disadvantage SSHelper is it only runs on non-rooted devices. Besides, SSHelper requires Android version 3.2, API 13, or newer, so SSHelper will not work on older Android devices.

As mentioned in the previous section, we also investigate on other protocols such as FTP, Samba. Through our experiments, we conclude that FTP is fast, but the permissions (owner, group, authorization) might not be transferred properly when the commands *wget* or *ftp* are used. Also for some reason the access to some folders is denied on a rooted device, using ftp, as listed in the following example, where the *lftp* [28] "mirror" command is used on a rooted Android target-device.

It seems to be rather difficult to make a connection with an SMB-service on a non-rooted Android target-device once an SMB-app is installed. It is likely that this might be caused by the SMB-service running on the Android target, which uses non-standard TCP-ports. Microsoft Windows and Linux operating systems only connect to SMB-servers running on conventional TCP-ports: 137-139. TCP-ports below 1024 are not usable on non-rooted Android devices because of security issues. Because of this limitation we decided do not use this protocol in our approach.

*B. Forensic implementations of installing apps on Android target*

In order to test this feature, an Android Emulator can be used. The Android Emulator and the Android Software Development Kit (SDK) are part of the Android software Development Toolkit (ADT) Bundle [29].The Android SDK provides developer tools necessary to build, test, and debug apps for the Android operating system. The set-up of such an environment is documented online, which we used to download, install and set-up the developer environment.

Using the "adb install" command the SSHDroid-package can be installed on the emulator. Before the installation a snapshot of the /data directory is made by using the "adb pull" command. After the installation again a snapshot is made. After uninstallation of SSHDroid, a third snapshot is made. By using the program Winmerge, the contents of both snapshot 1 and 2 are compared in order to see which artefacts would be visible after the installation related to the SSHDroid app. It showed that the app was installed in a newly created directory named:**"/data/data/berserker.android.apps.sshdroidpro"** Next, snapshot 2 and 3 were compared. This time only two files had changed: packages.list and packages.xml. The Android operating-system stores the names of the installed packages the file: packages.list. This method of package-listing can be compared to common Linux operating-systems like Debian GNU/Linux or Ubuntu where the Aptitude package-manager stores a list of installed packages in the file: /var/lib/aptitude/pkgstates. When an application is removed from the system, the entry, with the name of the package in it, is removed from this list, thus the package-file is modified.

*C. Testing environment*

In order to make the wireless connection, every basic Wi-Fi access-point can be used. It should however be verified that the access-point is NOT connected to the internet. A Wi-Fi network must be created on the access-point for the Android-devices to connect.

Note that the menu-settings of the Android target-device MUST be accessible in order to setup a Wi-Fi connection with the access-point. If the screen of the Android-device is locked, than it is not possible to set up the device for a file-transmission using Wi-Fi.

The Wi-Fi access-point uses DHCP to assign an IP-address to the target-device. For this reason the DHCP feature should be enabled on the access-point. We do not use Google Play app-store for downloading and installing the SSHDroid-package. Instead the investigating computer will run a web-server (Apache) to serve the SSH-application to the Android target-device. To do so, the investigating computer must be connected to one of the LAN ports on the access point. It can be advised to give the investigating computer a static IP. We should check that the access-point will provide an IP to the Android target-device in the same IP range as the investigating computer. So the DHCP scope on the access-point should be set-up correctly. On the investigating computer a Linux operating system is installed, running an Apache web-server and an OpenSSH-server. The SSHDroid-package will be served to the network using Apache-webserver. For example: the SSHDroid.apk package-file can be put in a shared web-directory named "pub" or "public": /var/www/pub on the investigator computer. It must be checked that the web-server is up and running and that the "public" share is accessible. If not, the webserver configuration should be adjusted accordingly. Instead of Apache another web-server may be used. The setup and configuration of a webserver and an access-point is beyond the scope of this paper.

For testing purposes we used the following Android target-devices:

- Samsung Apollo GT-I5800, Android version 2.2 (rooted)
- Archos 101G9 Tablet, Android version 4.0.4 (rooted)
- HTC One X, Android version 4.1.1 (non-rooted)

To install the package, we use the native web-browser on the Android target-device and point it to the web-address of the "public" share on the web-server onto which the SSHDroid package is served, for example: http://192.168.1.100/public/. A directory listing will be shown. We can download the SSHDroid package by tapping on the "SSHDroid.apk" link. Figure 1 shows the screenshot of Samsung Apollo, GT5800

Android device, browsing to the public directory on the webserver.

From the running SSHDroid-app on the Android target-device it will be easy to see which dynamic IP address is leased from the access-point's DHCP-server. Take note of this IP, because it is the IP that will be needed to connect to from the investigating computer.

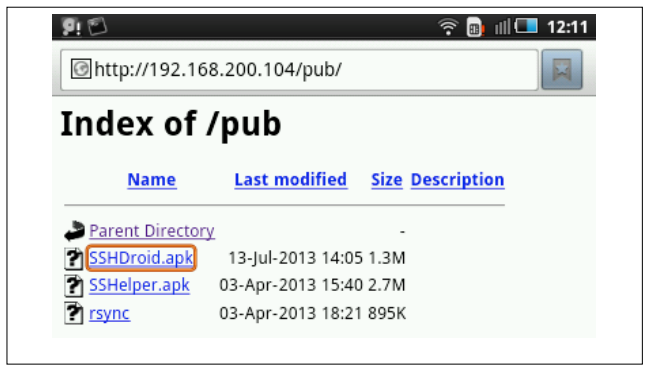

Fig. 1. Screenshot of browsing the public directory.

Figure 2 shows a running SSHDroid application on a non-rooted HTC One X Android device. As can be seen, SSHDroid assigned tcp-port: "2222" to the ssh-server instead of the common ssh port number: "22". This is known to the fact that non-rooted devices cannot connect locally to TCP ports below 1024

*D. Results and discussion*

As mentioned in the previous section, we use a script to facilitate the extraction process. Once the script is started it will ask for input, like the evidence-number, IP-address and SSH-port-number to connect, SSH-username and password, directory to back-up and the inclusion (not implemented yet in script) or exclusion of files and directories. The script will create a container-file using the Linux-command "*dd*". This file will be mounted and opened for reading. Next, *ssh* and *rsync* commands will be used to retrieve the data, which will directly put into this container. When the retrieval of data is finished the container-file will be un-mounted. When this has been succeeded the container file can be opened in most forensic applications for further examination of its contents.

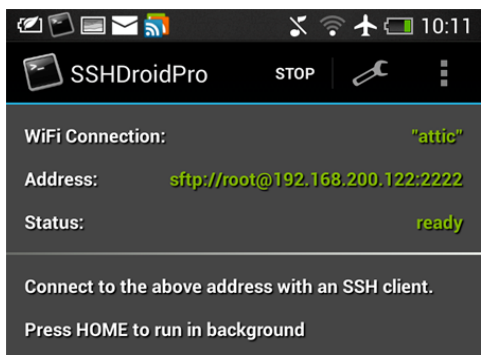

Fig. 2. Running SSHDroid application on a non-rooted HTC One X

Besides, the script must be executed as user "root". A "normal" user cannot run the script, because the loop-function, used in the script, needs elevated user-rights. For security reasons a non-root user will normally not have the privileges to execute a "mount" command using a loop-device. The "loop" functionality is needed to mount the *"dd"* container-file locally. The screenshot below (Fig. 3) shows a successful login to the SSHDroid ssh-server, which is running on the Archos G9 101 Android tablet.

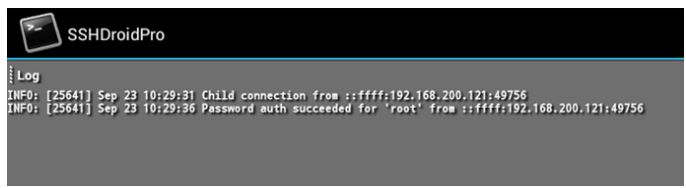

Fig. 3. Login to the SSHDroid ssh-server on Archos G9101

As mentioned in the Introduction section, in this paper, we also try to answer questions related to the feasibility of using Wi-Fi to extract data from Android devices. Through our experiments, we can now discuss on these questions:

- Is data extraction from an Android-device possible using Wi-Fi? In fact, we described an alternative method of fetching data from Android-devices in this paper. It showed that it is possible to make a connection to the Android file-system using Wi-Fi and retrieve data successfully.
- Can we extract all data as we do as with USB, like extracting deleted data? Wi-Fi data acquisition could be a valid alternative to an USB cable-connection for transfer of data. In both cases applies that the Android devices must be "rooted" to gain to all data. If the device is non-rooted, access will be permitted to several system-folders. In that case only the accessible files and folders will be preserved. Only on rooted Android devices and with using an USB-connection, it is possible to gain physical access to a memory block-device (for example: a memory-card). The *"dd"* command can be used with the *adb*-toolset to acquire a forensic image from an Android memory-device. Only using this last method it is possible to preserve deleted data. When a Wi-Fi connection is used for transferring data, we can only do the logical extraction of the data from the Android file-system, not physical one. Deleted data can never be secured or recovered using logical extraction. Only physical extraction can.
- Which network-protocols for Wi-Fi extraction can be used and which is best? For wireless communication using Wi-Fi, different protocols can be used. Research in this paper is focused on some of these protocols. The advantages and disadvantages of using these protocols were described above. Narrowing down the options, our recommendation is the usage of the *rsync* and *SSH* commands and protocols.

- Are there existing applications in the Google Play app-store that could be used for this and if so, which app could be used best? In fact, several apps were mentioned and tested. Finally the usage of the SSHDroid application, found in the Google Play app-store, is recommended for use in this paper.
- Will the file-system of the Android target-device remain untouched? It is clear that the file-system of the device during an examination will be changed. In order to communicate with the Android-device using Wi-Fi, an application must be installed onto the device. The app will be installed in the /data/app folder of the root file-system of the device. The impact of installing the SSHDroid-app on an Android target-device was also described above.
- Will the file attributes be preserved during file-transfer? When data from an Android target-device is retrieved using *rsync* with proper parameters, the file-attributes of the transferred data, like: timestamps and user-privileges, will be preserved.
- Will the given solution work on all versions of Android? In fact, we tested our approach with different version of Android: 2.2, 4.0 and 4.1. We can conclude that this solution can work on different versions of Android.

V. CONCLUSION

In this paper, we showed how Wi-Fi could be used to acquire data from an Android-device. In order to make a connection using Wi-Fi an application must be installed on the device. This application can be found in the Google Play app-store. Using a Linux-computer and a wireless access-point, data from an Android-device can be retrieved. The best results are obtained when *ssh* and *rsync* protocols are combined. Due to this research, we recommend to use the Android application SSHDroid. However, the SSHDroid-package does not contain the *rsync* executable. Therefore it might be useful in the future to write an SSH-server application to include this *rsync*-executable. We also write a bash-script that can be used as an example to show what could be done to automate the process of file-transfer using Linux bash scripting. The script can surely be improved, for example additional error checking and hashing-functionality (md5/sha1) could be added. To make it platform-independent it could be rewritten using program-languages like Python or Perl. Perhaps a GUI could be added. Existing tools could also be used to gain access to the Android file-system using SSH. For example tools with a graphical interface.

Besides, the downside of using existing tools is that it will not always be visible what the tools are doing. Will they touch the remote file-system and how? Will they preserve attributes of the retrieved data in a forensic manner? Using *rsync*, like shown in this paper, the process of file-transfer can be controlled into detail. When additional *rsync*-parameters are used and thoroughly tested, it is ensured that the output will be valid, matching the original file-system.